\documentclass[aps,prl,twocolumn,showpacs,showkeys,footinbib,longbibliography,superscriptaddress]{revtex4-1}

\usepackage{amsmath}
\usepackage{amssymb}
\usepackage{graphicx}
\usepackage{bm}
\usepackage{braket}
\usepackage{epstopdf}
\usepackage{orcidlink}

\usepackage{soul,xcolor,cancel} 

\begin{document}
\title{Microwave Signatures of Topological Superconductivity in Planar Josephson Junctions}
\author{Bar{\i}\c{s} Pekerten$^+$ \orcidlink{ 0000-0002-5794-9706}}
\email{barispek@buffalo.edu}
\affiliation{Department of Physics, University at Buffalo, State University of New York, Buffalo, New York 14260, USA}
\author{David Brand{\~a}o$^+$ \orcidlink{0000-0002-5772-1832}}
\email{dbrandao@buffalo.edu}
\thanks{$^+$These authors contributed equally to this work.}
\affiliation{Department of Physics, University at Buffalo, State University of New York, Buffalo, New York 14260, USA}
\author{Bassel Heiba Elfeky \orcidlink{0000-0003-3200-0040}}
\affiliation{Center for Quantum Information Physics, Department of Physics, New York University, New York 10003, USA}
\author{Tong Zhou \orcidlink{0000-0003-4588-5263}}
\affiliation{Department of Physics, University at Buffalo, State University of New York, Buffalo, New York 14260, USA}
\affiliation{Eastern Institute for Advanced Study, Eastern Institute of Technology, Ningbo, Zhejiang 315200, China}
\author{Jong E. Han \orcidlink{0000-0002-5518-2986}}
\affiliation{Department of Physics, University at Buffalo, State University of New York, Buffalo, New York 14260, USA}
\author{Javad Shabani \orcidlink{0000-0002-0812-2809}}
\affiliation{Center for Quantum Information Physics, Department of Physics, New York University, New York 10003, USA}
\author{Igor \v{Z}uti\'{c} \orcidlink{0000-0003-2485-226X}}
\email{zigor@buffalo.edu}
\affiliation{Department of Physics, University at Buffalo, State University of New York, Buffalo, New York 14260, USA}

\date{\today}
 
\begin{abstract}
Planar Josephson junctions provide a platform to host topological superconductivity which, through manipulating Majorana bound states (MBS), could enable fault-tolerant quantum computing. However, what constitutes experimental signatures of topological superconductivity and how MBS can be detected remains strongly debated. In addition to spurious effects that mimic MBS, there is a challenge to discern the inherent topological signals in realistic systems with many topologically-trivial Andreev bound states, determining the transport properties of Josephson junctions. Guided by the advances in microwave spectroscopy, we theoretically study Al/InAs-based planar Josephson junction embedded into a radio-frequency superconducting quantum interference device to identify microwave signatures of topological superconductivity. Remarkably, by exploring the closing and reopening of a topological gap, we show that even in a wide planar Josephson junction with many Andreev bound states, such a topological signature is distinguishable in the resonance frequency shift of a microwave drive and the ``half-slope'' feature of the microwave absorption spectrum. Our findings provide an important step towards experimental detection of non-Abelian statistics and implementing scalable topological quantum computing.  
\end{abstract}

\maketitle

The interest in topological superconductors reflects their fascinating fundamental properties and potential applications using manipulation of Majorana bound states  (MBS)~\cite{Kitaev2003:AP,Nayak2008:RMP,Alicea2011:NP,DasSarma2015:NPJQI,Matos-Abiague2017:SSC,Gungordu2022:JAP,Laubscher2021:JAP}. However, detecting such topological superconductivity and MBS remains a long-standing challenge~\cite{Amundsen2024:RMP,Yu2021:NP,Chen2019:PRL,Lee2012:PRL}. Revisiting Sr$_2$RuO$_4$, expected for decades to be $p$-wave spin-triplet which could support MBS~\cite{Mackenzie2003:RMP,Rice2004:S,Nelson2004:S,DasSarma2006:PRB}, or quantized zero-bias conductance peak (ZBCP)~\cite{Sengupta2001:PRB} as an experimental MBS signature~\cite{Mourik2012:S} both provide cautionary guidance for various difficulties in detection of superconducting properties~\cite{Zutic2005:PRL,Chen2019:PRL,DasSarma2021:PRB}. 
 
\begin{figure}[t]
\centering
\includegraphics*[width=\columnwidth]{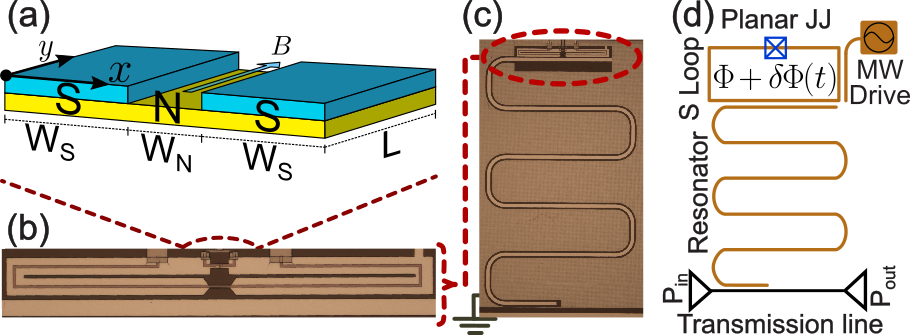}
\caption{(a) A planar JJ formed by two Al superconducting (S) regions covering InAs-based 2DEG normal region (N). Coordinate axes and JJ dimensions are indicated. Alternating current, $I$, flows along $\pm\hat{x}$; the applied magnetic field is ${\bm B}$. (b)-(d) Optical images of an example fabricated SQUID resonator, which we consider in our theoretical results. Gold: Al, dark brown: the etched mesa, and light brown: gate and a flux line. (c) Device diagram with resonator capacitively coupled to a transmission line and inductively to a superconducting loop with a flux $\Phi$ and a single JJ. (d) Schematic description from (c) of the JJ, SQUID, microwave drive line inducing fluctuations in $\Phi\rightarrow\Phi+\delta\Phi(t)$, resonator, and transmission line with input (output) power $P_\text{in}$ ($P_\text{out}$).}
\label{fig:System} 
\end{figure}

We focus on planar Josephson junctions (JJs)~\cite{Shabani2016:PRB,Hell2017:PRL,Pientka2017:PRX,Fornieri2019:N,Ren2019:N,Setiawan2019:PRB,Dartiailh2021:PRL,Banerjee2023:PRB}, where topological superconductivity is realized through proximity effects in the normal region (N)  of two-dimensional electron gas (2DEG), which separates superconducting (S) regions, shown in Fig.~\ref{fig:System}(a). Such JJs, with a phase or gate control and multi-terminal geometries suitable for probing non-Abelian statistics~\cite{Fu2008:PRL,Zhou2020:PRL,Zhou2022:NC}, could overcome limitations for topological superconductivity in 1D nanowires requiring fine-tuned parameters~\cite{Laubscher2021:JAP,Lutchyn2010:PRL,Oreg2010:PRL} and MBS detection through ZBCP. 

Our approach builds on advances in circuit quantum electrodynamics and highly-sensitive microwave spectroscopy of Andreev bound states (ABS)~\cite{Nazarov2003:PRL,Desposito2001:PRB,Bretheau2013:N,Reynoso2012:PRB,Tosi2019:PRX}. Due to electron-hole symmetry, ABS come in pairs and with energies $\pm E_\mathrm{ABS}(\phi)$, that depend on the relative superconducting phase, $\phi$, between the two S regions. In contrast, MBS are topologically-protected ABS with $E \approx 0$ localized to the ends of the junction that cannot be doubly occupied. To investigate these states, we theoretically consider an experimentally accurate model system where the planar JJ is embedded into a radio frequency superconducting quantum interference device (rf-SQUID), driven at a microwave frequency, $f_d$. The absorbed photons with $E=h f_d$ cause transitions between the ABS and change the JJ's critical current, $I_c$. By coupling the rf-SQUID to a microwave resonator, one can calculate this effect from the resulting change in the resonant frequency, $f_r$, of the resonator via a capacitively coupled transmission line~\cite{Park2009:PRL,Pozar2012,Elfeky2023:PRXQuantum}, as shown in Fig.~\ref{fig:System}. This approach, using numerical discretization, allows us to calculate the ABS spectrum and topological phase diagram of the planar JJ, as well as the microwave spectra and the shift in the critical current. 

Realistic planar JJs, in contrast with  many microwave studies focusing on a few ABS which is a characteristic for nanowires, atomic contacts, or constrictions, can have hundreds of topologically-trivial ABS, which complicate detecting inherent properties of topological superconductivity and MBS~\cite{Deacon2010:PRL,Zhang2022:PRL,Janvier2015:S,Wallraff2004:N,Tosi2019:PRX, vanWoerkom2017:NP, Hays2021:S,Pekker2013:PRL,Matute-Canadas2022:PRL}. Based on the experimental parameters of our fabricated JJs, we theoretically identify microwave signatures for closing and reopening of a topological gap. Our calculations show that such signatures are distinguishable in the transitions between the ABS due to a microwave drive of frequency, $f_d$, which change the current-phase relation (CPR), $I(\phi) = \Sigma_n I_n^\mathrm{ABS}$ that depends on the occupation of the ABS levels. This change in the CPR results in a shift in $f_r$, measured in the microwave circuit~\cite{Park2009:PRL,Pozar2012,Elfeky2023:PRXQuantum}.

In microwave detection, a quality factor $\gtrsim 10,000$ and a linewidth of $0.1-1\;$MHz yield fast ($\sim\mu$s) and neV energy resolution of ABS~\cite{Tosi2019:PRX,Hays2018:PRL}. In contrast to ZBCP measurements, this high sensitivity and  fermion-parity conservation, which enables the probing of quasiparticle poisoning and dynamical properties of superconducting devices~\cite{Kurilovich2021:PRB,Kos2013:PRB,Hays2018:PRL,Elfeky2023:PRXQuantum}, make our findings important beyond signatures of topological superconductivity and MBS. These findings could also advance the microwave studies in many other systems with a large number of ABS, for example, in superconducting spintronics~\cite{Linder2015:NP,Eschrig2015:RPP,Cai2023:AQT, Amundsen2024:RMP}, in nonreciprocal phenomena such as diode effects~\cite{Amundsen2024:RMP,Baumgartner2022:NN,Dartiailh2021:PRL, Pekerten2022:PRB,Lotfizadeh2024:CP}, or in emerging superconducting qubits~\cite{Krantz2019:APR,Tafuri:2019,Pita-Vidal2020:PRA}. 

We consider a model Al/InAs-based JJ with $L = 4\;\mu$m and $W_N = 80\;$nm [see Fig.~\ref{fig:System}], embedded in an rf-SQUID, that is inductively coupled to a hanger $\lambda/4$ coplanar waveguide resonator for our calculations. The model resonator is capacitively coupled to a transmission feed line, allowing for the complex transmission and the resonant response to be measured. We use $f_r\sim 7\;$GHz, estimated from experimentally relevant circuit parameters~\cite{Park2009:PRL,Pozar2012}. We therefore present our numerical results with $f_d$ up to a few $f_r$ ($\sim 40$ GHz), unless showing the full range of $h\,f_d \lesssim 2\Delta_0\sim 120$ GHz is instructive.

Our model for the microwave-driven JJ is described by the Bogoliubov-de Gennes (BdG) Hamiltonian~\cite{Pientka2017:PRX, Zhou2022:NC} 
\begin{align}
H_0 & = \left[\frac{\mathbf{p}^2}{2m^\ast} - \mu\left(x,y\right) + \frac{\alpha}{\hbar}\left(p_x\sigma_y - p_y\sigma_x \right)\right]\tau_z\nonumber\\
	&\quad - \frac{g^\ast\mu_B}{2}\mathbf{B}\cdot\boldsymbol{\sigma} + \Delta\left(x,y\right)\tau_+ + \Delta^\ast\left(x,y\right)\tau_-\;,
\label{eq:H0}
\end{align}
for which numerically solve the corresponding discretized eigenvalue problem~\cite{SM_MW_1} to obtain ABS. Here, ${\bf p}$ is the momentum, $\mu\left(x,y\right)$ is the chemical potential, $\alpha$ is the Rashba spin-orbit coupling (SOC) strength, $\mathbf{B} = B\,\hat{y}$ is the in-plane magnetic field and $\mu_B$ is the Bohr magneton, while $m^\ast$ and $g^\ast$  are the effective mass and $g$-factor, respectively. The proximity-induced superconductivity in the InAs 2DEG is described by the pair potential $\Delta(x,y) = \Delta_0\,\Theta(|x-W_N/2|)\,\exp{(i\,\phi(x, y)/2})$, where $\Theta$ is the  step function and in $\phi(x, y) = \phi_0 \, \mathrm{sgn}{(x)}$, $\phi_0$ is uniform in each S region. Similarly, we take $\Delta_0\rightarrow \Delta_0(B, T)$ to be real and uniform in the S regions, having the usual BCS dependence on $B$ and temperature $T$ \cite{Tinkham:1996, Zhou2022:NC}. $\sigma_i$ ($\tau_i$) are Pauli (Nambu) matrices in the spin (particle-hole) space, and $\tau_\pm=(\tau_x\pm i\tau_y)/2.$ 

\begin{figure}[t]
\centering
\includegraphics*[width=\columnwidth]{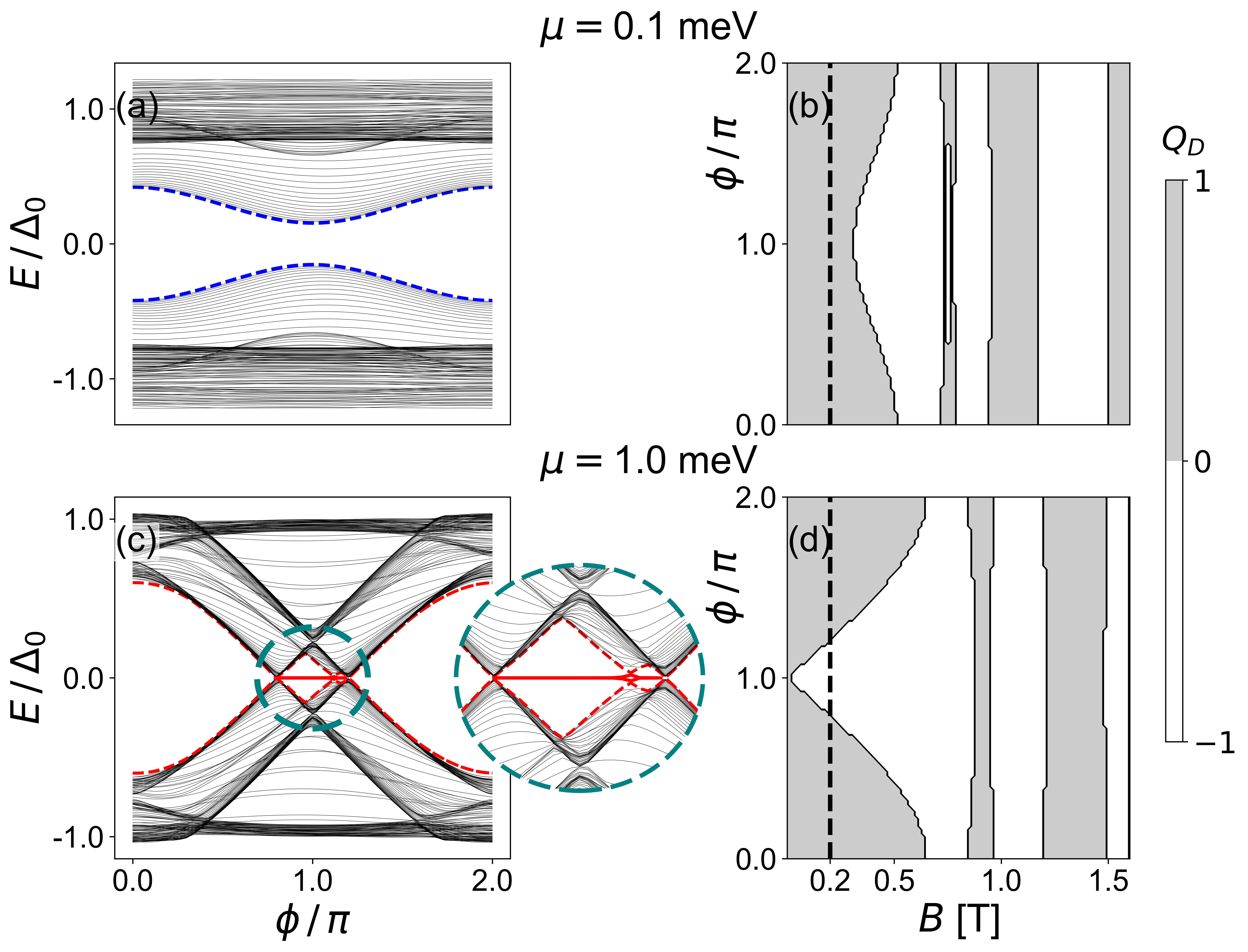}
\caption{(a), (c) ABS spectra as a function of $\phi$ for JJ  from Fig.~\ref{fig:System}(a). $W_S=600\;$nm, $W_N=80\;$nm, $L=4\;\mu$m, and $B=0.2\;$T. (b), (d) The corresponding topological charge, $Q_D$, as a function of $B$ and $\phi$. (a), (b) For $\mu=0.1\;$meV, the gapped ABS are trivial for all $\phi$.  (c) For $\mu=1.0\;$meV, there is a gap closing and reopening. The inset: MBS at $E\approx 0$ (red solid line). (d) A topological transition as $\phi$ is varied at $B=0.2\;$T, corresponding to the gap closing points in (c).}
\label{fig:Spectra} 
\end{figure}

\begin{figure*}[ht]
\centering
\includegraphics*[width=\textwidth]{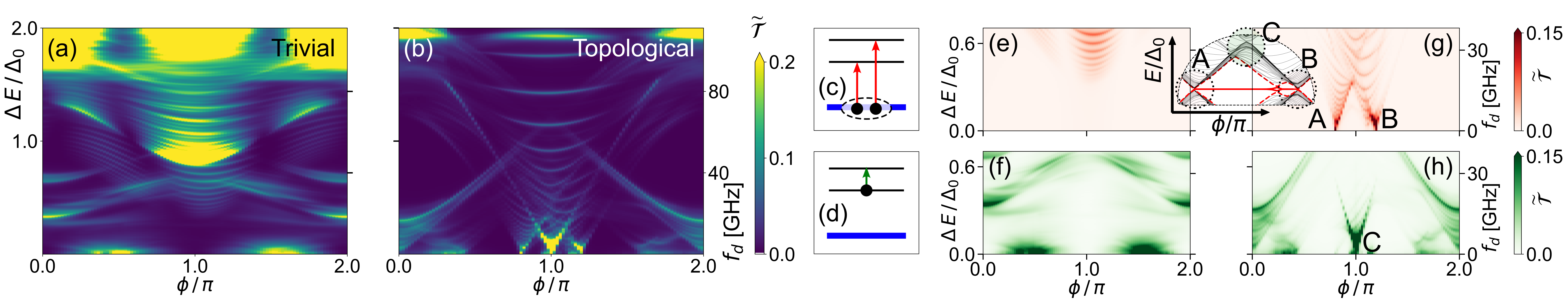}
\caption{(a), (b) Transition lines $\widetilde{\mathcal{T}}(f, \phi)$, calculated using Eq.~(\ref{eq:Trans}), corresponding to $hf_d=\Delta E = |E_n-E_m|$ between ABS from Figs.~\ref{fig:Spectra}(a) and (c), with a Lorentzian broadening parameter $\epsilon=30\;$mK. In the schematic of pair (c) and single-particle (d) transitions where the ground state is denoted by a blue line, the sum of the arrow lengths in (c) gives $\Delta E$. Resolving (e) pair and (f) single-particle transitions from (a) up to $40\;$GHz shows no gap closing and opening. In contrast a similarly resolved (g) pair and (h) single-particle transitions from (b) reveal signatures of topological superconductivity. The inset: an enlarged part of the Fig.~\ref{fig:Spectra}(c) spectra. The transitions in the circles ``A'' and ``B'' of the inset lead to the \textsf{\textbf{V}}-shaped features ``A'' and ``B'' in (g), signaling a gap closing and reopening. Transitions in the circle ``C'' lead to the \textsf{\textbf{V}}-shaped feature ``C'' in (h).}
\label{fig:Transitions} 
\end{figure*}

We choose material parameters consistent with our fabricated epitaxial Al/InAs JJs which support robust proximity-induced superconductivity and topological transitions~\cite{Dartiailh2021:PRL}: $\Delta_0=0.23\;$meV, $m^\ast=0.027 m_0$ where $m_0$ is the  electron mass, $g^\ast=10$ for InAs, $\alpha=10$ meVnm, critical magnetic field $B_c(T=0)=1.6\;$T and temperature $T=30\;$mK.  The corresponding ABS spectra are shown in Fig.~\ref{fig:Spectra}. In the top row, for $\mu=0.1\;$meV and $B=0.2\;$T, the JJ is in the topologically trivial phase for all $\phi$. In Fig.~\ref{fig:Spectra}(a) there is no gap closing at $B=0.2\;$T, while in Fig.~\ref{fig:Spectra}(b) the calculated topological charge~\cite{Nayak2008:RMP}, $Q_D=1$, confirms this trivial phase.The mirror symmetry in the system with respect to the middle of the junction along the $x$-direction allows for an effective time reversal symmetry, leading to a system that can support MBS in a wider range of $\phi$ values around $\phi=\pi$, as seen in Figure~\ref{fig:Spectra}(b) and (d)~\cite{Pientka2017:PRX}. For a finite $W_S$, there are normal reflections and an imperfect transparency at the N/S interface, causing the topological phase diagram to deviate from the typical ``diamond shape''~\cite{Pientka2017:PRX, Pekerten2022:PRB, Pekerten2024b:PRB} with a trivial phase even at $\phi=\pi$ and small $B$. In contrast, for a larger $\mu$, the JJ in our model undergoes topological transitions even for $B\sim 0$ at $\phi=\pi$, leading to a wider range of $\phi$ that can support MBS for higher $B$ [Fig.~\ref{fig:Spectra}(d)]. At $B=0.2\;$T, MBS near $E\sim 0$ is seen in the spectrum in Fig.~\ref{fig:Spectra}(c) and the inset for $0.8\pi<\phi<1.2\pi$.~\ref{fig:Spectra}~\ref{fig:Spectra} In Supplemental Material (SM)~\cite{SM_MW_1}, we complement these results with CPR, topological gap, the MBS localization, as well as topological phase transitions for larger $\mu$ as $B$ is varied. 

We next discuss the signatures of the topological gap closing and reopening in Fig.~\ref{fig:Spectra}(c) using the microwave spectroscopy with the setup from Figs.~\ref{fig:System}(b)-(d). The microwave drive induces a flux fluctuation, $\Phi(t) = \Phi(0) + \delta\Phi(t)$, in the rf-SQUID around its unperturbed value $\Phi(0)$, which leads to the JJ phase fluctuation $\delta \phi(t)$, described by the perturbed Hamiltonian~\cite{Desposito2001:PRB,Zazunov2003:PRL,Janvier2015:S, Hays2021:S,Olivares2014:PRB,Vayrynen2015:PRB}
\begin{align}
H(\phi + \delta \phi(t)) &= H_0(\phi) + \delta \phi(t) \, H_\mathrm{MW},
\label{eq:H0MW}
\end{align}
where the perturbation $H_\mathrm{MW} = \partial H_0(\phi)\, / \, \partial \phi$ is given by
\begin{align}
H_\mathrm{MW} = -i\,(\Delta_0/2)\,\left(e^{-i\phi(\mathbf{r})/2}\,\tau_+ - e^{i \phi(\mathbf{r})/2}\,\tau_-\right).
\label{eq:HMW}
\end{align}
For our model experimental setup based on fabricated JJs embedded in microwave resonator circuits, we can examine the validity of this approach by a simple estimate, $\delta \phi/2\pi \sim M \, I_\mathrm{MW} /\Phi_0 < M \, I_c /\Phi_0 \approx 0.05$, where we use an experimentaly relevant value of $M\sim 120\;$pH for the mutual inductance and a critical current of $I_c\lesssim 1\;\mu$A as an upper bound for the microwave driving current $I_\mathrm{MW}$~\cite{Dartiailh2021:NatCommun, Phan2023:PRA}, thus justifying the treatment of the effects of the microwave drive as a perturbation~\cite{Vayrynen2015:PRB,Wesdorp2024:PRB,Chidambaram2022:PRR}. The transitions caused by this perturbation conserve the fermion parity~\cite{Bretheau2017:NP,Janvier2015:S,Kos2013:PRB,Vayrynen2015:PRB,SM_MW_1}, which reduces the number of the allowed transition lines shown in Figs.~\ref{fig:Transitions}(a) and (b). We can then distinguish the pair transitions, between a Cooper pair and two ABS, or the single-particle transitions between two ABS, as denoted in Figs.~\ref{fig:Transitions}(c) and (d). In the BdG Hamiltonian, pair (single-particle) transitions correspond to transitions between the states with opposite (same) energy sign. 

With high energy- and time-resolution, microwave spectroscopy provides information about the allowed transitions lines, $h\, f_d^{nm} = |E_m-E_n|$, with the transition probability $|\mathcal{M}_{nm}|^2$ governed by the matrix element $\mathcal{M}_{nm} = \braket{\psi_m|H_\mathrm{MW}|\psi_n}$, where $\psi_n$ is the eigenstate with eigenenergy $E_n$ of $H_0$ [Eq.~(\ref{eq:H0})]. Our calculated results for planar JJs in Fig.~\ref{fig:Transitions} show the allowed transition lines
\begin{align}
\mathcal{T}(f_d, \phi) =\sum_{mn} |\mathcal{M}_{nm}|^2 \mathcal{F}(E_m, E_n)\, \mathcal{L}(E_{mn}-hf_d), 
\label{eq:Trans}
\end{align}
where $\mathcal{F}(E_m, E_n)$ is the thermal occupation factor and $\mathcal{L}(E) = (\epsilon/\pi)/\left(E^2+\epsilon^2 \right)$ the Lorentzian broadening, parameterized by $\epsilon$. Figures~\ref{fig:Transitions}(a) and (b), with the transition lines normalized to their maximum value, $\widetilde{\mathcal{T}} =\mathcal{T}/\mathcal{T}_\mathrm{max}$, correspond to the system described by Figs.~\ref{fig:Spectra}(a) and (c), respectively.  While Fig.~\ref{fig:Transitions}(a) which clearly shows the bulk bandgap, as we see a very limited transition probability around $E \approx 0$ and $\phi \approx \pi$, we can conclude that Fig.~\ref{fig:Transitions}(b) displays the gap closing signature of topological superconductivity. 

Unlike the majority of the microwave studies with a much smaller number of ABS~\cite{Janvier2015:S,Wallraff2004:N,Tosi2019:PRX,vanWoerkom2017:NP,Hays2021:S} with theoretical results that we can recover by reducing the system size of our model JJs~\cite{Bretheau2017:NP,Dartiailh2021:NatCommun,Elfeky2023:ACSN}, the several hundred ABS considered in Fig.~\ref{fig:Transitions} complicate the identification of topological superconductivity. To address this, we resolve the calculated transition lines into pair and single-particle transitions, now replotted up to $f_d= 40\;$GHz, compatible with relevant experimental setups for microwave detection, instead of the full ABS energy range of $2\Delta_0$ or $\sim 120\;$GHz.

By comparing these two types of transitions for the trivial [Figs.~\ref{fig:Transitions}(e) and (f)] and topological phase [Figs.~\ref{fig:Transitions}(g) and (h)], we see some striking differences. This is further illustrated in the inset of Fig.~\ref{fig:Transitions}(g), which is the enlarged version of Fig.~\ref{fig:Spectra}(c) focusing on low-energy ABS spectra near the topological gap closing and reopening, and the matching features in Fig.~\ref{fig:Transitions}(g) and (h). The features marked by ``A'' and ``B'' in the inset denote the locations of topological gap closing and reopening and the MBS at $E\approx 0$ between these points. The matching \textsf{\textbf{V}}-shaped features seen in Fig.~\ref{fig:Transitions}(g), also marked by ``A'' and ``B,'' resolve the pair transition lines around the gap closing and reopening, which therefore mark the microwave signatures of the topological superconductivity. On the other hand, the \textsf{\textbf{V}}-shaped single-particle transition lines seen in Fig.~\ref{fig:Transitions}(h), marked by ``C'', correspond to the transitions between the ABS levels in a region far from $E\approx 0$ state (marked by ``C'' in the inset) and therefore are {\em not} related to topological superconductivity. We also note that the features near zero frequency at $\phi\sim\pi/2, 3\pi/2$ in Fig.~\ref{fig:Transitions}(a), (b), (f) and (h) are related to single-particle transitions between ABS near $E\sim\Delta_0$, and are not topology related.

While, to the best of our knowledge, these here identified microwave signatures of topological superconductivity were not previously discussed, it is reassuring that for the trivial phase, we recover the behavior known from the previous JJ studies. For example, by reducing the number of ABS that we consider, we retain all the main features of the previous work on the weak-link JJs~\cite{Tosi2019:PRX}. This also includes the observed band gap around $\phi=\pi$ in Figs.~\ref{fig:Transitions}(e) and (f), as expected for the trivial phase.

\begin{figure}[t!]
\centering
\includegraphics*[width=\columnwidth]{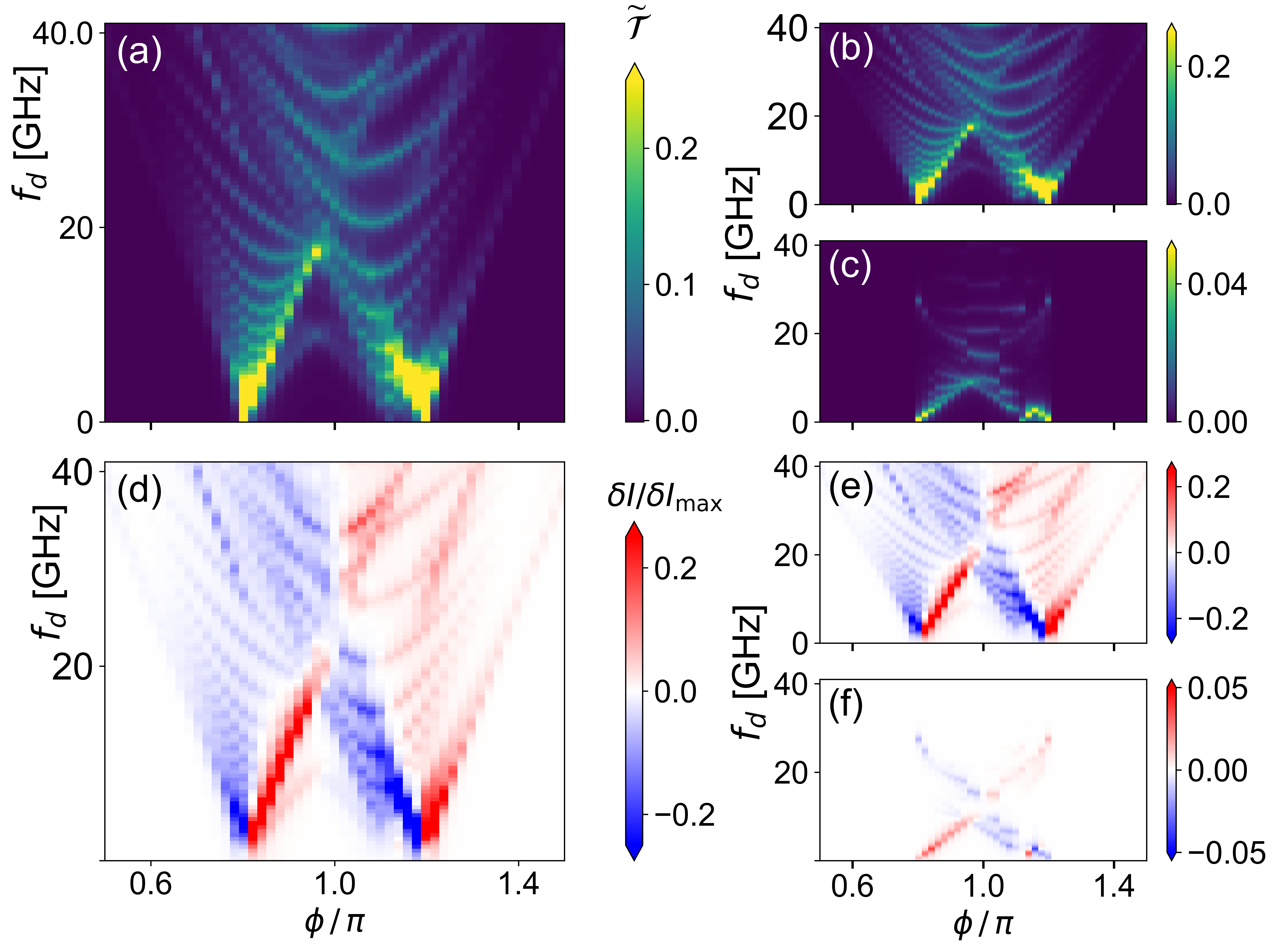}
\caption{(a)-(c) The transition lines and (d)-(f) $\delta I(f, \phi)$ for the ABS from Fig. 2(c) with suppressed higher-energy single-particle transitions. The lines  $\widetilde{\mathcal{T}}(f, \phi)$ in (a) are resolved into (b) pair and (c) single-particle transitions. The remaining transition lines in (c) are due to the energy-split overlapping MBS. (d) $\delta I(f, \phi)$  displays the gap closing and reopening. $\delta I(f, \phi)$ shown from (e) pair and (f) single-particle transitions, reveals the gap closing and opening as well as the transitions from $E\approx0$ split MBS.}
\label{fig:Shift} 
\end{figure}

To provide further guidance to experimental work and previous $T=0$ results for a small number of Andreev bound states ~\cite{Blais2021:RMP,Bretheau2014:PRB,Olivares2014:PRB}, we consider the effects of the microwave drive on the JJ's CPR, which affects the $f_r$ as discussed above. Without the microwave drive, the free energy of the JJ is $F=-k_B T \sum_{E_n} \mathrm{ln}\left[2\,\mathrm{cosh} \left( E_n/2 k_B T \right) \right]$~\cite{Tinkham:1996}, where $k_B$ is the Boltzmann constant and $\{E_n\}$ is the spectrum for the JJ. The CPR, $I(\phi) = (2e/\hbar) \, \partial F\,/\, \partial\phi$, is then 
\begin{align}
I(\phi) &= -\frac{e}{\hbar}\, \sum_{E_n}\,\frac{\partial E_n}{\partial \phi} \, \mathrm{tanh}\left( \frac{E_n}{2 k_B T} \right).
\label{eq:CPR} 
\end{align}
Equation~(\ref{eq:CPR}) shows the contribution of each energy level to the supercurrent, $(e/\hbar)\, \partial E_n/\partial\phi$, weighted by a thermal occupation $\mathrm{tanh}(E_n/2 k_B T)$. We consider a change to $I(\phi)$ in Eq.~(\ref{eq:CPR}) as a function of $f_d$, due to the microwave drive. At short timescales, we phenomenologically express the change in $I(\phi)$ as
\begin{align}
\delta I(f, \phi) 	&= \Gamma\,\sum_{m, n}\, \left(\frac{\partial E_m}{\partial\phi}-\frac{\partial E_n}{\partial\phi}\right)\,|\mathcal{M}_{mn}|^2\nonumber\\
			&\quad \times\, \mathcal{F}(E_m, E_n)\, \mathcal{L}(E_{mn}-hf_d), 
\label{eq:deltaI}
\end{align}

We choose a special form of the thermal occupation $\mathcal{F}(E_m, E_n) = f_\mathrm{FD} (E_n) \, \left(1-f_\mathrm{FD}(E_m) \right),$ where  $f_\mathrm{FD}$ is the Fermi-Dirac function, to achieve the effect of suppressing single-particle transitions between the levels away from $E\approx 0$, due to the doubling of the fermionic degrees of freedom in the BdG Hamiltonian. This form of $\mathcal{F}(E_m, E_n)$ allows us to simulate the suppression of quasiparticle transitions between energy levels $E_{m,n}$ away from 0. The identification of topological signatures is simplified in the resulting transition lines $\widetilde{\mathcal{T}}$ and the shift in the current $\delta I(f,\phi)$. A counterpart plot with a conventional thermal occupation factor is discussed in SM~\cite{SM_MW_1}. In the limit of $T=0\;$K and $\epsilon=0$, our Eq.~(\ref{eq:deltaI}) recovers the same form as in the published literature~\cite{Blais2021:RMP, Bretheau2014:PRB, Olivares2014:PRB}. While Eq.~(\ref{eq:deltaI}) does not include the $\phi$-dependent change of the slope and curvature of the ABS, our key results and trends are preserved by this simplification~\cite{Park2009:PRL,SM_MW_1}.

\begin{figure}[t!]
\includegraphics[width=\columnwidth]{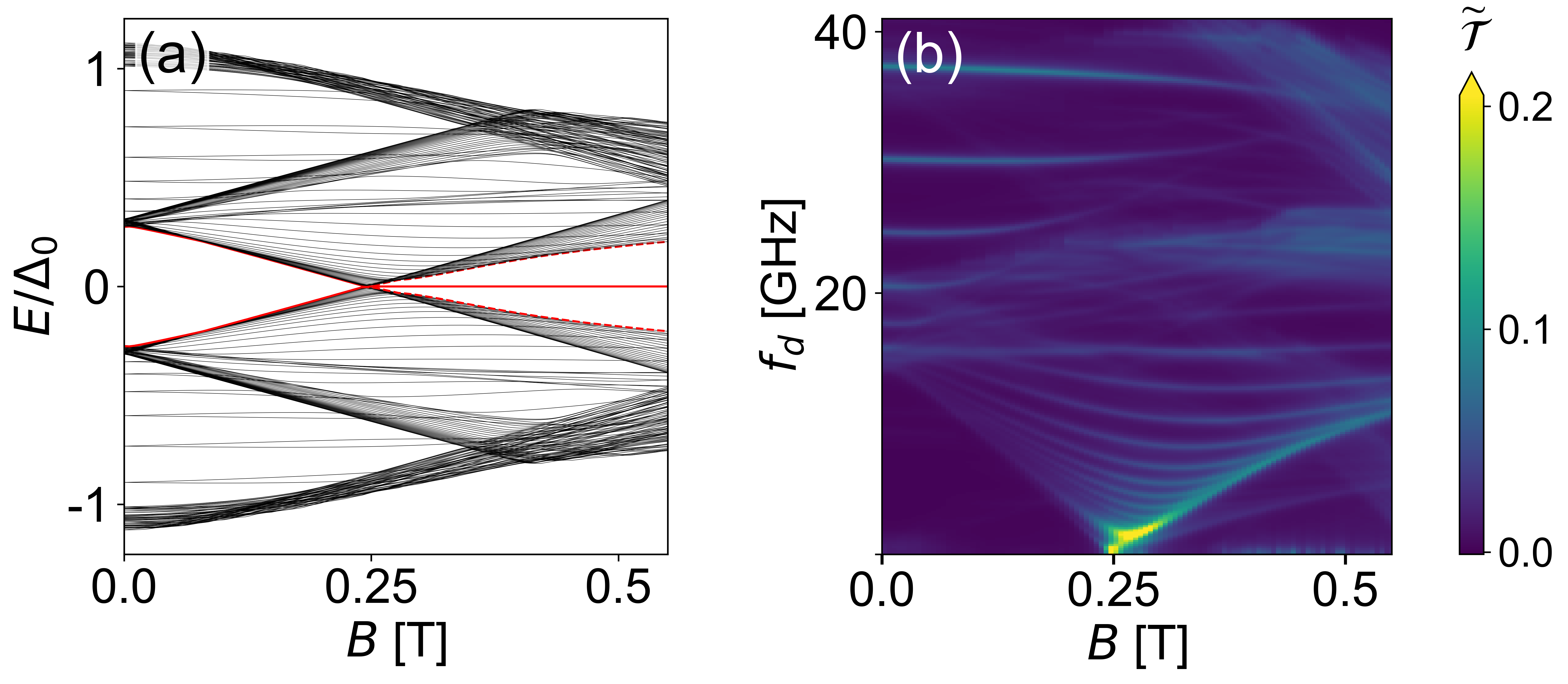}
\caption{(a) ABS spectrum as a function of $B$ showing the lowest 240 levels with $\mu=1$ meV and $\phi=0.75\pi$. The gap closing and reopening, and the appearance of the zero-energy MBS at $B=0.24\;$T match the topological phase transition (see SM~\cite{SM_MW_1}). (b) Transition lines $\widetilde{\mathcal{T}}$ as a function of $B$. \textsf{\textbf{V}} feature indicating that the gap is closing and reopening is visible at $B=0.24\;$T near $f_d\sim 0$.}
\label{fig:MW_Zeeman}
\end{figure}

\begin{figure}[t!]
\includegraphics[width=\columnwidth]{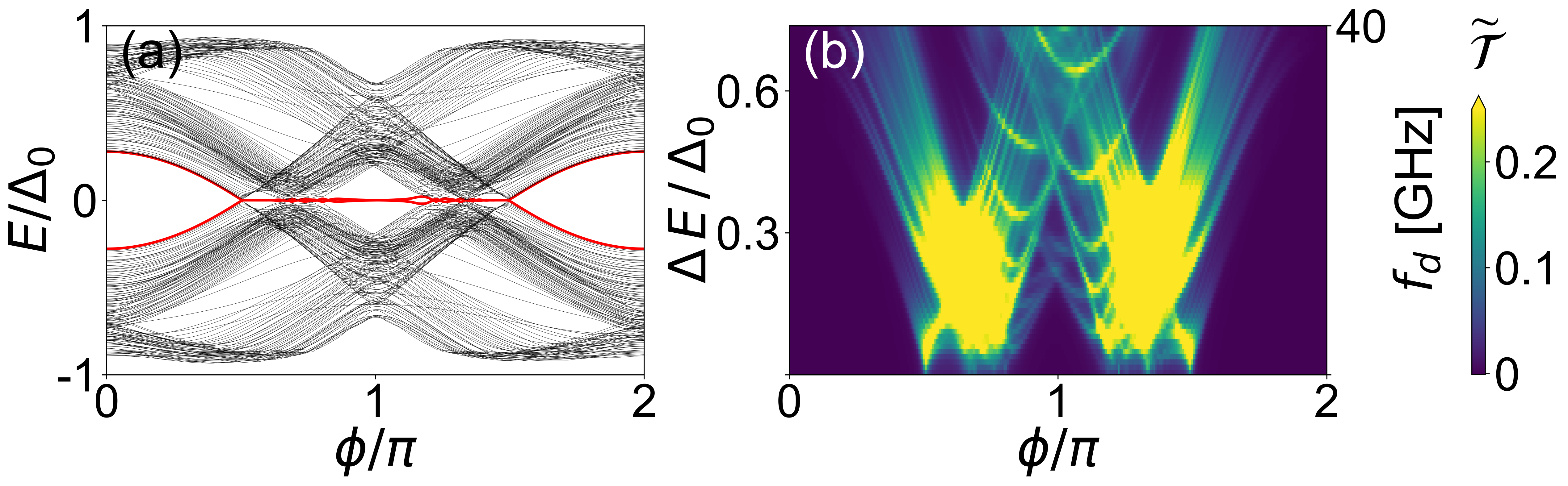}
\caption{(a) ABS spectrum as a function of $\phi$ showing the lowest 240 levels with $\mu=5$ meV and $B=0.5\;$T, for $L=4\;\mu$m. (b) Transition lines $\widetilde{\mathcal{T}}$ as a function of $B$. The double \textsf{\textbf{V}} feature, indicating the gap closing and reopening, is less sharp but visible around $\phi=0.75\pi$ and $\phi=1.25\pi$ near $f_d\sim 0$.}
\label{fig:LargeMu}
\end{figure}

For the JJ from Fig.~\ref{fig:Spectra}(c), with single-particle transitions where $|E| \gtrsim k_B T$, we use this approach to plot the transition lines in Figs.~\ref{fig:Shift}(a)-(c) and $\delta I(f, \phi)$ of Eq.~(\ref{eq:deltaI}) in Figs.~\ref{fig:Shift}(d)-(f). Both the transition lines and $\delta I(f, \phi)$ reveal the signatures of the gap closing and opening, even for a system with hundreds of ABS. While these low-frequency transitions near the gap closing points appear in Figs.~\ref{fig:Shift}(b) and (e) for the pair transition, an even more striking feature becomes visible in Figs.~\ref{fig:Shift}(c) and (f) for the single-particle transitions. The suppression of the higher-energy single-particle transitions, including the feature ``C'' in Fig.~\ref{fig:Transitions}(h), removes the background that competes with the signal from transitions between the energy-split MBS and higher energies, now observed prominently in Fig.~\ref{fig:Shift}(c) as a transition line between $\phi\sim 0.8\pi$ and $\phi\sim 1.2\pi$ with half of the slope of the \textsf{\textbf{V}} shapes marked by ``A'' and ``C'' in Fig.~\ref{fig:Transitions}(g)~\cite{Vayrynen2015:PRB, Murthy2020:PRB}. This ``half-slope'' feature is also observed in Figs.~\ref{fig:Transitions}--\ref{fig:LargeMu} in the panels that depict topological transitions. The time-resolved capabilities of microwave spectroscopy could then also provide a detection of the non-Abelian statistics in planar JJs~\cite{Zhou2020:PRL,Zhou2022:NC}, which is supported by experimentally measured parameters, but is very challenging using conventional transport measurements. 

Our overall features in Fig.~\ref{fig:Shift} are consistent with other work on ABS in quantum dots~\cite{Deacon2010:PRL, Zhang2022:PRL}, constrictions~\cite{Janvier2015:S,Wallraff2004:N} and nanowires~\cite{Tosi2019:PRX, vanWoerkom2017:NP, Hays2021:S}. This is important for our effort to expand the microwave spectroscopy as a sensitive probe for topological superconductivity in systems with a large number of ABS, where additional signatures become more pronounced by resolving pair and single-particle transitions. For example, one could apply our proposed detection to topological JJs based on 2D materials~\cite{Xie2023:PRL}, or systems with a dominant cubic SOC where the topological superconductivity is expected to be $f$-wave~\cite{Alidoust2021:PRB,Luethi2022:PRB}. It would be interesting to test if the enhanced signatures, obtained by removing the single-particle transitions (violating fermion parity conservation), could be implemented experimentally by identification and dynamical selection of parity in JJs, demonstrated in atomic contacts~\cite{vanWoerkom2017:NP,Bretheau2013:PRX} and weak links~\cite{Ackermann2023:PRB,Wesdorp2024:PRB}.

The \textsf{\textbf{V}} feature of topological gap closing and reopening is also seen in Fig.~\ref{fig:MW_Zeeman} from our numerical results for an experimental system where the JJ is phase biased and the Zeeman field is varied. We demonstrate the case where the phase bias is $\phi=0.75\pi$ (at $\phi=\pi$ the transition would be very close to $B=0\;$T). The ABS spectrum in Fig.~\ref{fig:MW_Zeeman}(a) and the topological phase diagram in SM~\cite{SM_MW_1} indicate a transition at $B=0.24\;$T. A corresponding \textsf{\textbf{V}} feature in the transition lines is observed in Fig.~\ref{fig:MW_Zeeman}(b). The resolved pair transitions, and the $\delta\,I(f, B)$ plots in SM~\cite{SM_MW_1} also demonstrate the \textsf{\textbf{V}} feature and the gap reopening, indicating that topological transitions may be observed in phase-biased microwave experiments.

We next examine the influence of a higher chemical potential in a JJ with many ABS levels. The corresponding calculated results are shown in Fig.~\ref{fig:LargeMu}. By comparing the results for the spectra in Figs.~\ref{fig:Spectra}(c) and \ref{fig:LargeMu}(a), we see that at a higher $\mu$ will be less defined at a constant $B$ as a function of $\phi$, since a wider range of ABS levels will cross $E=0$. This will result in a reduced topological gap and less protected MBS, and will lead to a less sharp \textsf{\textbf{V}}-shaped signature in the transition lines [Fig.~\ref{fig:LargeMu}(b)]. A larger $B$ will shift the gap closing points further apart by shifting the ABS lines in the spectra, making the gap reopening signature more distinct. Similarly, a larger finite $L$ will open the topological gap further, having the same effect of making the signatures more distinct (see SM~\cite{SM_MW_1}). In our simulations with $\mu=5\;$meV and $L=4\;\mu$m, $\alpha=10\;$meV nm at $B=0.5\;$T, the gap reopening signature remains prominently visible but less sharp in the transition lines.

Our findings have various implications beyond topological superconductivity. Microwave spectroscopy of ABS could provide a sensitive probe of nonreciprocal phenomena  in JJs~\cite{Amundsen2024:RMP}, while its temporal resolution could elucidate novel driving mechanism with time-dependent gate-control of SOC~\cite{Monroe2022:PRA}. As a result,  even in the absence of any bias current in the JJ, it is possible to consider enhanced opportunities for superconducting electronics and neuromorphic computing using voltage pulses corresponding to the single flux quantum~\cite{Tafuri:2019,Holmes2013:IEEETAS,Monroe2022:PRA,Likharev1991:IEEETAS,Crotty2010:PRE}. With decades of transport studies in junctions with only a single superconductor~\cite{Kashiwaya2000:RPP}, microwave detection also offers a complementary approach to revisit various manifestations of ABS and unconventional superconductivity. 

\textit{Note added--}Recently, we were made aware of a publication~\cite{Trif2018:PRB} titled ``Current susceptibility as a probe of Majorana bound states in nanowire-based Josephson junctions,'' which is relevant to our work.

\acknowledgments
We thank Prof. Leonid Glazman for valuable discussions. This work is supported by the U.S. ONR MURI Award No. N000142212764 and U.S. ONR Award No. N000141712793. Computational resources were provided by the UB Center for Computational Research.  

%

\end{document}


\title{Supplemental Material: \\
Microwave Signatures of Topological Superconductivity in Planar Josephson Junctions}
\author{Bar{\i}\c{s} Pekerten$^+$}
\email{barispek@buffalo.edu}
\affiliation{Department of Physics, University at Buffalo, State University of New York, Buffalo, New York 14260, USA}
\author{David Brand{\~a}o$^+$}
\email{dbrandao@buffalo.edu}
\thanks{$^+$These authors contributed equally to this work.}
\affiliation{Department of Physics, University at Buffalo, State University of New York, Buffalo, New York 14260, USA}
\author{Bassel Heiba Elfeky}
\affiliation{Center for Quantum Information Physics, Department of Physics, New York University, New York 10003, USA}
\author{Tong Zhou}
\affiliation{Department of Physics, University at Buffalo, State University of New York, Buffalo, New York 14260, USA}
\affiliation{Eastern Institute for Advanced Study, Eastern Institute of Technology, Ningbo, Zhejiang 315200, China}
\author{Jong E. Han}
\affiliation{Department of Physics, University at Buffalo, State University of New York, Buffalo, New York 14260, USA}
\author{Javad Shabani}
\affiliation{Center for Quantum Information Physics, Department of Physics, New York University, New York 10003, USA}
\author{Igor \v{Z}uti\'c}
\email{zigor@buffalo.edu}
\affiliation{Department of Physics, University at Buffalo, State University of New York, Buffalo, New York 14260, USA}

\maketitle

\date{\today}
 
\section{Tight-Binding Calculations}\label{SECT:SM:TB}

The discretized tight-binding (TB) calculations  for the epitaxial Al/InAs-based planar Josephson junctions (JJs)  throughout this work are carried out using the Kwant package \cite{Groth2014:NJP}. The materials parameters and the JJ dimensions are specified in the main text. We first formulate the Bogoliubov-de Gennes (BdG) Hamiltonian in Eqs.~(2) and (3) (in the main text) and their terms into their tight-binding form
\begin{align}
H_0^\mathrm{TB}	&= \sum_{i, j} H_0^\mathrm{onsite}\,\ket{\mathbf{r}_{i,j}}\bra{\mathbf{r}_{i,j}}  +\big(H_0^\mathrm{up}\,\ket{\mathbf{r}_{i,j+1}}\bra{\mathbf{r}_{i,j}} \nonumber\\
		         	& \quad  + H_0^\mathrm{right}\,\ket{\mathbf{r}_{i+1,j}}\bra{\mathbf{r}_{i,j}} + \mathrm{h.c.}\big),
\label{eq:SM:TBHamiltonian}
\end{align}
where h.c. represents a Hermitian conjugate and
\begin{align}
H_0^\mathrm{onsite}	&= [4t -\mu^\mathrm{TB}(\mathbf{r}_{i,j})]\,\tau_z + B^\mathrm{TB}\,\sigma_y \nonumber\\
					&+ \Delta(\mathbf{r}_{i,j}) \, \left(e^{i\,\phi (\mathbf{r}_{i,j})/2} \tau_+ + e^{-i\,\phi(\mathbf{r}_{i,j})/2} \tau_- \right)\, ,  \\
H_0^\mathrm{up}		&= -t \,\tau_z + \frac{\alpha^\mathrm{TB}}{2a}\,\sigma_x\tau_z\, , \\
H_0^\mathrm{right}	&= -t \,\tau_z - \frac{\alpha^\mathrm{TB}}{2a}\,\sigma_y\tau_z\, .
\label{eq:SM:TBHamiltonian_Parts}
\end{align}
Here, $a$ is the TB lattice parameter, $t=\hbar^2/(2 m^\ast a^2)$ is the hopping parameter, $m^\ast$ is the effective mass, $i$ and $j$ are the integer indices running over the $x$- and $y$-coordinates of the TB lattice:  $i\in [-(W_S+W_N/2)/a,\, (W_S+W_N/2)/a]$ and $j\in [0, \,L/a]$), where $W_S$  and $W_N$ are the widths of the superconducting (S) and the normal region (N) of the JJ, while $L$ is its length.  $\mathbf{r} \rightarrow \mathbf{r}_{i,j} = ia\,\hat{x}+ja\,\hat{y}$ are the discretized coordinates of the TB lattice. We consider a uniform chemical potential $\mu^\mathrm{TB}(\mathbf{r}_{i,j}) = \mu(\mathbf{r}) = \mu$, applied magnetic field,  $B^\mathrm{TB}=B$, and the strength of the Rashba spin-orbit coupling, $\alpha^\mathrm{TB} = \alpha$, which have the same definitions as in Eq.~(2). $\sigma_i$ ($\tau_i$) are Pauli (Nambu) matrices in the spin (particle-hole) space, and $\tau_\pm=(\tau_x\pm i\tau_y)/2.$ The pair potential $\Delta(\mathbf{r}_{i,j}) = \Delta_0 \,\Theta(|i-L/2a|)$ is nonzero and uniform in the superconducting (S) regions,  with real $\Delta_0$, and the superconducting phase difference is $\phi(\mathbf{r}_{i,j}) = \phi_0\, \mathrm{sgn}(i) \,\Theta(|i-W_N/2a|)$, where $\Theta$ is the step function. Finally, the TB form of the microwave perturbation Hamiltonian in Eq.~(4) is 
\begin{align}\label{eq:SM:TB_MWHamiltonian}
H^\mathrm{TB}_\mathrm{MW}	&= \frac{i}{2} \sum_{i, j} \Delta(\mathbf{r}_{i,j}) \, \mathrm{sgn}(i)\,\ket{\mathbf{r}_{i,j}}\bra{\mathbf{r}_{i,j}} \nonumber\\
							& \times \, \left( e^{i\,\phi(\mathbf{r}_{i,j})/2} \tau_+ -  e^{i\,\phi(\mathbf{r}_{i,j})/2} \tau_- \right).
\end{align}

The spectra in Fig.~2 (of the main text), the eigenfunctions, including those plotted in Fig.~S\ref{fig:SM:CPR}(e), and the matrix elements $\mathcal{M}_{mn}$, giving the transition between the Andreev bound states (ABS),  used in Figs.~3 and 4 are calculated by diagonalizing the TB Hamiltonians in Eqs.~(\ref{eq:SM:TBHamiltonian}) and (\ref{eq:SM:TB_MWHamiltonian}). The eigenvalues, $E_n$, are calculated by diagonalizing $H_0^\mathrm{TB}$ from the spectra plotted in Fig.~2. We consider the lowest $N=240$ eigenvalues, chosen such that all the eigenvalues inside the superconducting gap, $E_n \in (-\Delta_0, \Delta_0)$ are included for all $\phi$. Microwave spectroscopy provides information about the allowed transitions lines, $h\, f_d^{nm} = |E_m-E_n|$,  where $f_d$ is the microwave drive frequency, with the transition probability $|\mathcal{M}_{nm}|^2$, governed by the matrix elements $\mathcal{M}_{mn} = \braket{\psi_m|H^\mathrm{TB}_\mathrm{MW}|\psi_n}$, which are calculated using the eigenvectors $\psi_n$ corresponding to the eigenvalues $E_n$. 

For the planar JJ, to calculate the topological charge, $Q_D$, in Fig.~2(d) and the current-phase relation (CPR)  in Fig.~S\ref{fig:SM:CPR}, we utilize a translationally-invariant form of the Hamiltonian $H_0^\mathrm{TB}$ in the $y$-direction (along the N/S interface) with the wave number $k_y$ being a good quantum number. The translationally-invariant version of $H_0^\mathrm{TB}$ in Eq.~(\ref{eq:SM:TBHamiltonian}) is
\begin{align}\label{eq:SM:TB_TransInvHamiltonian}
H_0^\mathrm{TB}(k_y)		&= \left( e^{-i\,k_y a}\,H_0^\mathrm{down} + H^\mathrm{layer} + H_0^\mathrm{up} \, e^{i\,k_y a} \right)\,,  
\nonumber\\
H_0^\mathrm{TB}(k_y)\,\psi 	&= E(k_y) \, \psi\, ,
\end{align}
where $ H^\mathrm{layer} = H_0^\mathrm{onsite} + H_0^\mathrm{right} + H_0^\mathrm{left}$, $H_0^\mathrm{down} = (H_0^\mathrm{up})^\dagger$ and $H_0^\mathrm{left} = (H_0^\mathrm{right})^\dagger$. The topological charge is given by~\cite{Tewari2012:PRL}
\begin{align}
Q_D &= \frac{\textrm{Pf}\left(H(k_y=\pi/a)\right)}{\textrm{Pf}\left(H(k_y=0)\right)},
\label{eq:SM:Topocharge_TB}
\end{align}
where $\textrm{Pf}$ is the Pfaffian and its value $-1$ $(1)$ corresponds to the topological (trivial) state~\cite{Nayak2008:RMP}. For a translationally-invariant system, we take the spectrum in the free energy $\{E_n\}$ as $E(k_y)$, with $k_y$ chosen within the half-Brillouin zone $k_y \in [0, \pi/a]$. The ground state phase, $\phi_\mathrm{GS}$,is the superconducting phase difference that minimizes the free energy~\cite{Pientka2017:PRX}, which also corresponds to one of the zeroes of the CPR, $I(\phi)$~\cite{Pekerten2022:PRB}. $\pi$-jumps in $\phi_\mathrm{GS}$ approximately, but not necessarily exactly, correspond to topological phase transitions~\cite{Pientka2017:PRX,Pekerten2022:PRB,Dartiailh2021:PRL}.

\section{Topological Gap and Current-Phase Relation}
\label{SECT:SM:CPR_QD}

\begin{figure}[th]
\includegraphics[width=\columnwidth]{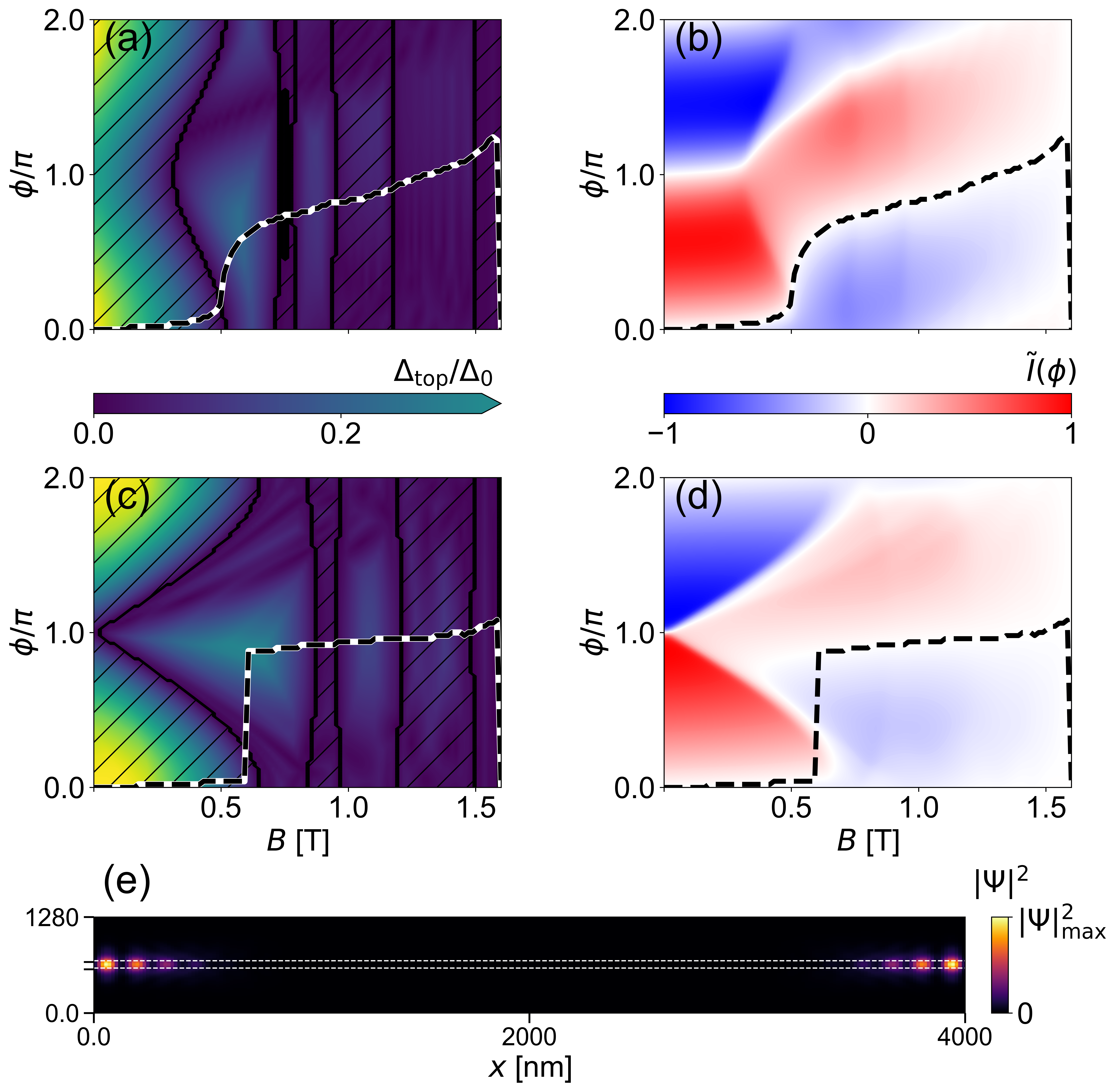}
\caption{Topological gap, $\Delta_\mathrm{top}$, and the current-phase relation (CPR), $\tilde{I}(\phi)$. (a), (b) [(c), (d)] for the JJ from Figs.~2(a), (b) [Figs.~2(c), (d)] with the chemical potential $\mu=0.1$~meV ($\mu=1$~meV). In (a), (c) trivial regions are dashed. In (a)-(d) dashed lines show the evolution of the ground-state phase, $\phi_\mathrm{GS}(B)$, with magnetic field. (e) The MBS probability for $\phi=\pi$ for the JJ in (c), (d). Dotted lines:  N/S boundaries.} 
\label{fig:SM:CPR}
\end{figure}

We show in Fig.~S\ref{fig:SM:CPR} the topological gap, $\Delta_\mathrm{top}$,  and the normalized current phase relation CPR, $\tilde{I}(\phi) = I(\phi)/|I_\mathrm{max}|$, for the system as a function of $B$ from Fig.~2. Specifically, Figs.~S\ref{fig:SM:CPR}(a) and (b) [Figs.~S\ref{fig:SM:CPR}(c) and (d)] correspond to the spectra from Fig.~2 (a)-(b) [Fig.~2(c)-(d)]. The dashed lines show $\phi_\mathrm{GS}$, unperturbed by the microwave drive. As discussed in Ref.~\onlinecite{Pekerten2022:PRB}, the $\phi$-jumps accompanying the transition to the topological phase need not be abrupt and $\Delta_\mathrm{top}$ can be small,  even for the system with $Q_D=-1$. Figure~S\ref{fig:SM:CPR}(e) shows the calculated probability density for the Majorana bound states (MBS), for $\mu=1$~meV and $B=0.2$~T, corresponding to Fig.~2(c) at $\phi=\pi$, where the system is topologically nontrivial. The horizontal dotted lines in Fig.~S\ref{fig:SM:CPR}(e) denote the N/S boundaries. The MBS localization at the JJ edges is clearly seen.

\section{Microwave Perturbation of the Critical Current}

\begin{figure}[h]
\includegraphics[width=\columnwidth]{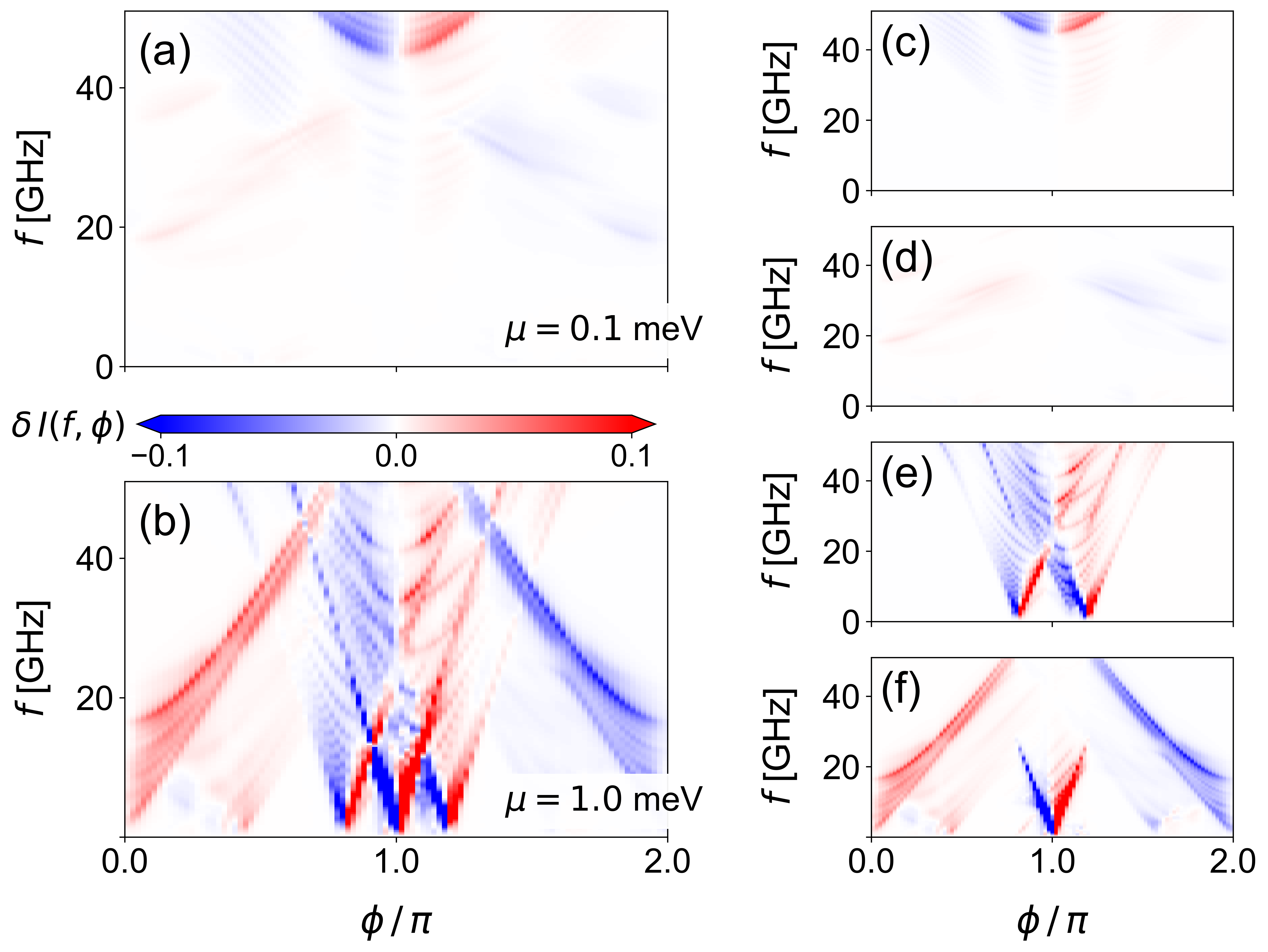}
\caption{TB calculations to model the CPR changes, $\delta I$, due to microwave transitions as a function of $\phi$ and microwave frequency, $f$, at a fixed $B=0.2$~T, corresponding to the ABS spectra in Fig.~2. The complete results in (a) [(b)] for trivial (topological) JJs are resolved into pair transitions in (c) [(e)] and single-particle transitions in (d) [(f)]. These results are the counterparts of Fig.~4, but include the influence of the quasiparticle poisoning on the single-particle transitions.}
\label{fig:SM:changecurrent}
\end{figure}

In this section, we revisit Fig.~4 and calculate again  the CPR changes, $\delta I(f, \phi)$, of Eq.~(5) in the main text, due to microwave transitions as a function of $\phi$ and microwave frequency $f$. However, we use a thermal occupation factor $\mathcal{F}$ that does not suppress the single particle transitions. Figures~S\ref{fig:SM:changecurrent}(a)-(c) show $\delta I(f, \phi)$ for the topologically-trivial case at all $\phi$. As expected, both the complete results for $\delta I$ ([Fig.~S\ref{fig:SM:changecurrent}(a)] and its contributions resolved into the pair [Fig.~S\ref{fig:SM:changecurrent}(c)] and single-particle transitions [Fig.~S\ref{fig:SM:changecurrent}(d)] show a fully gapped state. In contrast, Fig.~S\ref{fig:SM:changecurrent}(b), which show the complete results for $\delta I$ in the topological state, and  its contributions resolved into the pair [Fig.~S\ref{fig:SM:changecurrent}(e)] and single-particle transitions [Fig.~S\ref{fig:SM:changecurrent}(f)], reveal characteristic ``V" features, noted also in the Fig.~4.  However,  the central ``V'' around $\phi-\pi$ at low frequencies is no longer suppressed and is prominent. The cause of the features are explained in the main text in the discussion of Fig.~4.

\section{Signatures of the B-Field-Induced Topological Transition} 

\begin{figure}[htb]
\includegraphics[width=\columnwidth]{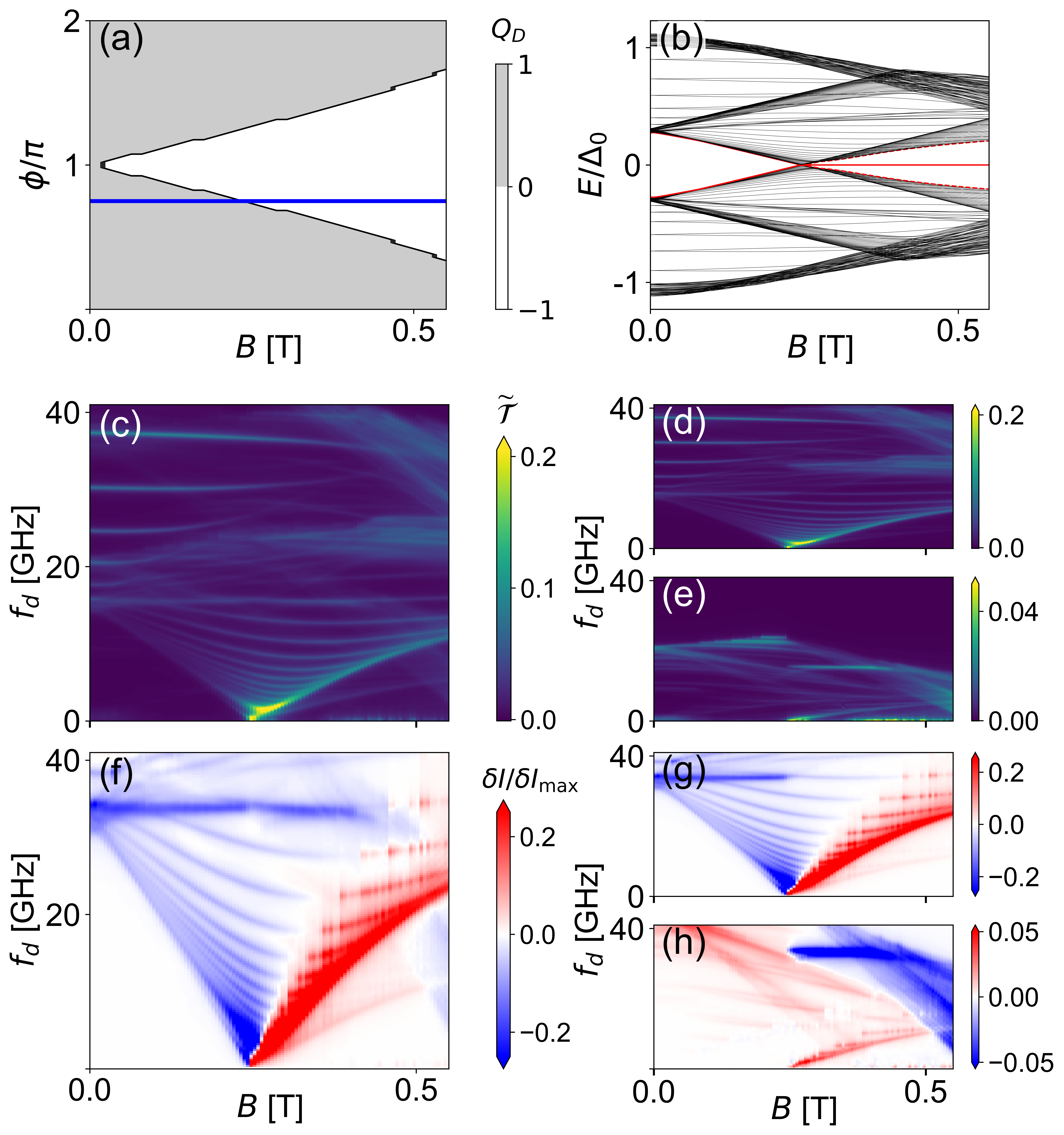}
\caption{(a) The topological phase diagram defined by the topological charge, $Q_D$, for the JJ as a function of $B$ and $\phi$ with $\mu=1$ meV, as in Fig.~S\ref{fig:SM:CPR}(c). The horizontal line marks $\phi=0.75\pi$, at which Figs.~S\ref{fig:SM:MW_Zeeman}(b)-(h) are plotted. (b) Low-energy JJ spectra consisting of 240 lowest energy levels. Absorption lines $\widetilde{\mathcal{T}}$, (c) their complete contribution, resolved into the (d) pair and (e) single-particle transitions as a function of $B$. ``V'' feature around $B=0.3$~T is visible. The evolution of $\delta I(f_d)$ as a function of $B$, (f) its complete contribution and resolved into the (g) pair and (h) single-particle transitions. The effect of $\Delta_\mathrm{top}$ closing is visible.}
\label{fig:SM:MW_Zeeman}
\end{figure}

We now consider the $B$-field-induced topological transitions~\cite{Pientka2017:PRX, Dartiailh2021:PRL, Pekerten2022:PRB} and their low-frequency microwave signatures. It is instructive to show the topological phase diagram, described by the corresponding $Q_D(\phi, B)$, for a system whose parameters coincide with those in Figs.~2 (c), (d) and Figs.~S\ref{fig:SM:CPR}(c), (d). We take $\phi=0.75\pi$ fixed and vary $B$ from 0~T to $0.6$~T, where we observe a topological transition around $B=0.3$~T. The topological transition does not need to be immediately followed by a reopening of a large $\Delta_\mathrm{top}$~\cite{Pekerten2022:PRB}. This is evident in the plot of the ABS spectra in Fig.~S\ref{fig:SM:MW_Zeeman}(b). The MBS presence is clearly seen (red lines) after the first gap closing and reopening around $B=0.3$~T. Near the next gap closing and reopening, around $B=0.9$~T (not shown), there is only a small $\Delta_\mathrm{top}$ and not fully protected and split MBS. As $B$ increases, there is a fanlike spectra. 

The gap closings from the spectra in Fig~S\ref{fig:SM:MW_Zeeman}(b) are also seen as the signature in the microwave transitions in Fig.~S\ref{fig:SM:MW_Zeeman}(c), even with a small $\Delta_\mathrm{top}$. Recalling the previously noted sharp ``V'' feature,  it can be seen in the first gap closing around $B=0.3$~T. The low-energy features with increasing $B$ correspond to the transitions from split MBS to higher ABS. As in Fig.~4, we confirm that with microwave spectroscopy we can observe the effects of gap closing and reopening as well as the corresponding transitions from the MBS. 

\section{Single-Particle Transitions with Suppressed Quasiparticle Poisoning}
\label{sec:SM:suppressedSPT}

\begin{figure}
\includegraphics[width=\columnwidth]{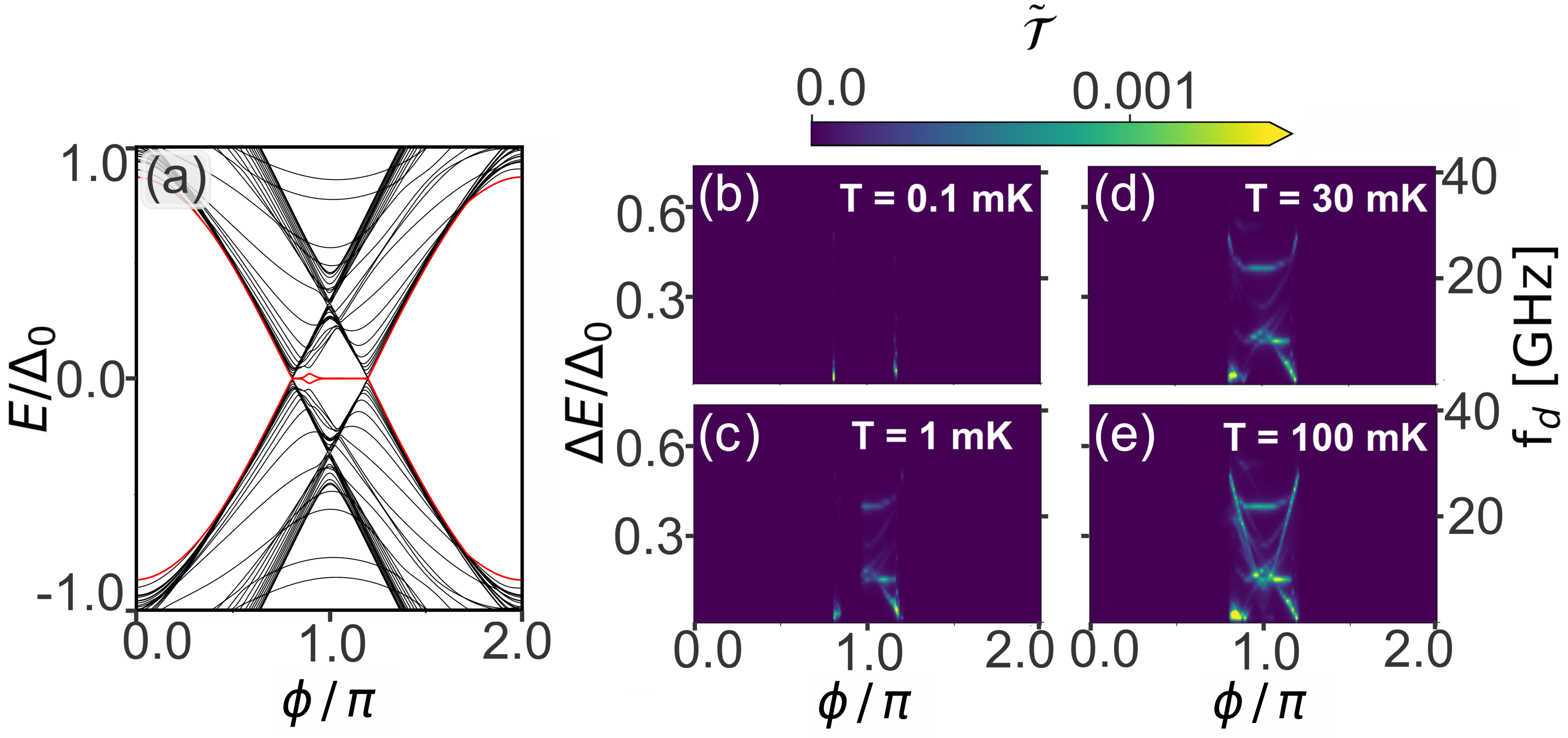}[t]
\caption{(a) Calculated ABS spectra as a function of $\phi$. (b)-(d) The transition lines, $\widetilde{\mathcal{T}}$, for the single-particle contributions, with suppressed quasiparticle poisoning, as a function of $\phi$, for different temperatures, $T$. In a finite planar JJ with dimensions $W_S=600$ nm, $W_N=80$ nm, $L=2 \,\mu$m, at $B=0.2$~T and $\mu=1$~meV, the zero-mode excitation in the topological region may exhibits a small gap, so that the states with $E>\mu$ can be occupied by thermal contribution.}
\label{fig:SM:ThermalSP}
\end{figure}

In Fig.~4, the transition lines, $\widetilde{\mathcal{T}}$,  as defined by Eq.~(5) in the main text, and the change in the CPR were discussed when the quasiparticle poisoning is suppressed. Specifically, Fig.~4(c) and (f) show the results for single-particle transitions, where the visible lines are due to transitions from the states $E\approx0$, which are thermally occupied. Due to the doubling of the fermion degrees of freedom, the thermal occupation factors for the BdG levels with negative energy should be treated as that of a hole occupation (lack of electrons) as opposed to the thermal occupation factor of an electron. However, ignoring this issue and treating all ABS levels in the BdG picture as electron levels has the effect of suppressing the single-quasiparticle transitions (transitions between ABS levels of the same energy sign) between ABS levels of energies not close to zero. The spectroscopic measurement time is order or orders of magnitude shorter than the quasiparticle lifetime of ~µs~\cite{Elfeky2023:PRXQuantum}, leading to a low probability of introducing quasiparticle poisoning into the system during the measurement. We use this form of the thermal occupation factor, to simulate a system with low quasiparticle poisoning, in Figs.~4 and 6 and in Figs.~S\ref{fig:SM:ThermalSP} and S\ref{fig:SM:ThermalSP_highmu}. 

All of the plots in the main text and in SM, except for Fig.~S\ref{fig:SM:ThermalSP}, are calculated at $T=30\;$mK, a representative temperature of the substrate temperature in experiments. For comparison, in Fig.~S\ref{fig:SM:ThermalSP}(a), the ABS spectra are given for a JJ with a smaller length, $L=2\,\mu$m (instead of $L=4\,\mu$m in the main text), and $\mu = 1$~meV, as a function of $\phi$. In Figs.~S\ref{fig:SM:ThermalSP}(b)-(e) the transition lines are shown for the single-particle contribution, with suppressed quasiparticle poisoning at different temperatures, $T=0.1\;$mK for Fig.~S\ref{fig:SM:ThermalSP}(b), $T=1\;$mK for Fig.~S\ref{fig:SM:ThermalSP}(c), $T=30\;$mK for Fig.~S\ref{fig:SM:ThermalSP}(d) and $T=100\;$mK for Fig.~S\ref{fig:SM:ThermalSP}(e). We see that a higher temperature leads to a higher transition amplitude, which is responsible for the brighter lines in Figs.~S\ref{fig:SM:ThermalSP}(b)-(e).

\section{Microwave-Induced Transitions for a Larger ${\bm \mu}$}

\begin{figure}[t]
\includegraphics[width=\columnwidth]{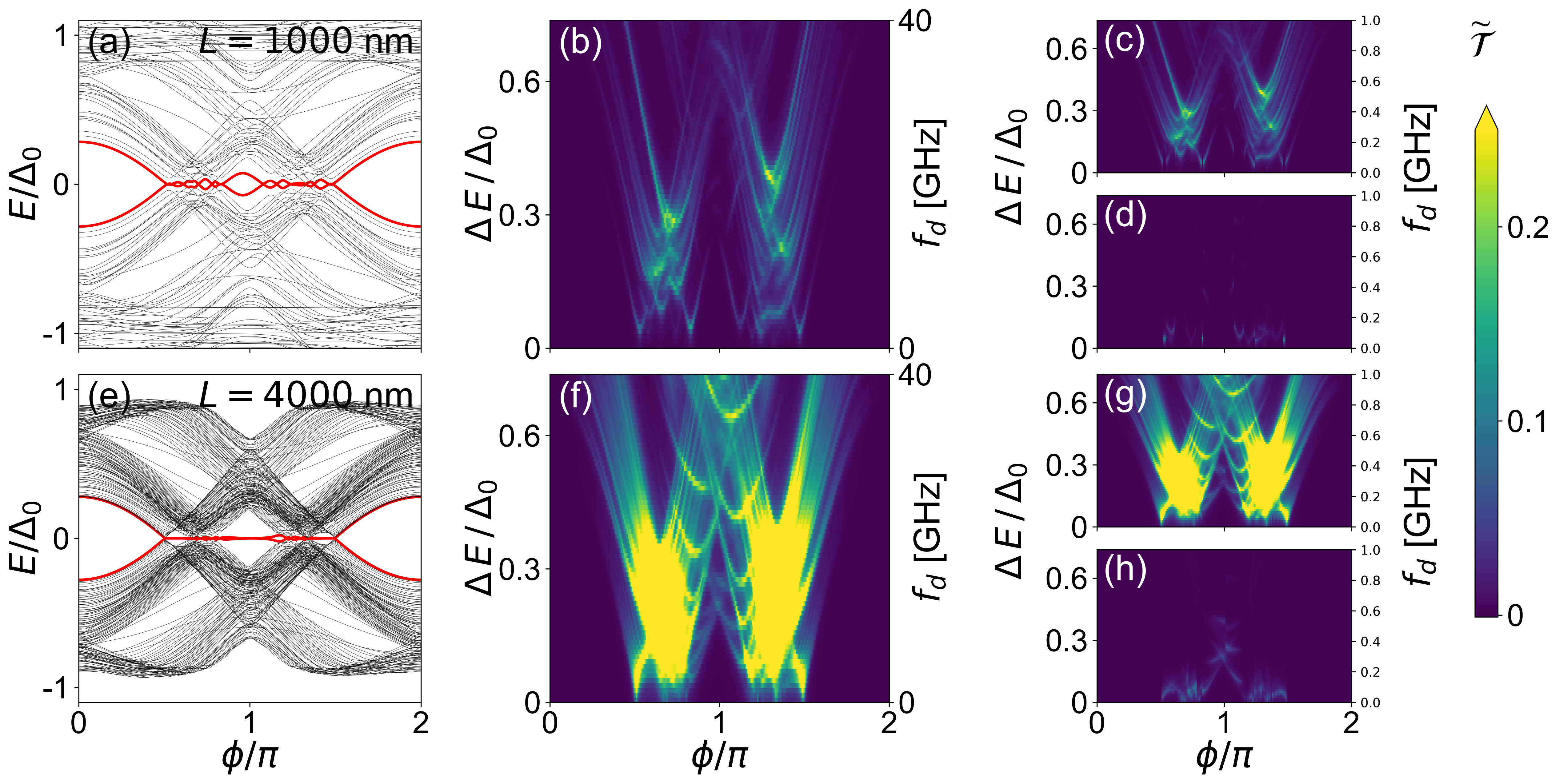}
\caption{Calculated ABS spectra and transition lines as a function of $\phi$ for $\mu = 5$~meV and $B=0.5$~T. (a) [(e)] ABS and (b)-(d) [(f)-(h)] transition lines spectra for $L=1\;\mu$m [$L=4\;\mu$m]. (b), (f) The full transition lines are calculated with a thermal factor suppressing single-particle transitions between energy levels away from $E=0$, as well as shown as resolved into (c), (g) pair and (d), (h) single-particle transitions.}
\label{fig:SM:ThermalSP_highmu}
\end{figure}

We finally consider the case of a larger $\mu=5$meV, where at smaller energies, many more ABS become relevant, as seen in Fig.~S\ref{fig:SM:ThermalSP_highmu}(a) and (e). Compared to the spectra for $\mu=1$~meV in Fig.~2(c), the topological region is larger along $\phi$, while sizable portions of the topological region have a nearly vanishing $\Delta_\mathrm{top}$ and the MBS (shown in red) are split in energy. In Figs.~S\ref{fig:SM:ThermalSP_highmu}(b)-(d) [(f)-(h)] we show the transition lines for this system, calculated using Eq.~(5) from the main text. This is expected as different ABS levels cross $E=0$ at different $\phi$ values, leading to a less well-protected MBS. Higher Zeeman field has the effect of shifting these levels further, this bunching them together slightly more, leading to a slightly larger region with sizable topological gap and a slightly sharper transition feature. Furthermore, a larger $L$ means the MBS are better separated and better protected, leading to a higher topological gap. We examine this by numerically plotting the ABS spectra for $L=1\;\mu$m [Fig.~S\ref{fig:SM:ThermalSP_highmu}(a)] and for $L=4\;\mu$m [Fig.~S\ref{fig:SM:ThermalSP_highmu}(e)]. While the topological region and the extent of the MBS in $\phi$ is the same, the difference in the topological gap, and the splitting of the MBS is evident. This is reflected in the full absorption spectrum [Fig.~S\ref{fig:SM:ThermalSP_highmu}(b) and S\ref{fig:SM:ThermalSP_highmu}(f)] and in the single particle and pair transition resolved absorption spectrum [Fig.~S\ref{fig:SM:ThermalSP_highmu}(c), S\ref{fig:SM:ThermalSP_highmu}(d), S\ref{fig:SM:ThermalSP_highmu}(g) and S\ref{fig:SM:ThermalSP_highmu}(h)]. The features of the topological transitions in the absorption spectrum are not as sharp, but they are still very distinguishable, both in the plotted complete transition lines [Fig.~S\ref{fig:SM:ThermalSP_highmu}(b)] as well as in their resolved contributions from the pair [Fig.~S\ref{fig:SM:ThermalSP_highmu}(c) and (g)] transitions.

%